\documentclass[aps,twocolumn,floats,prd,nofootinbib,superscriptaddress,showpacs,showkeys]{revtex4}
\usepackage[dvips]{graphicx} %
\usepackage{epsfig,amsmath}
\usepackage{amssymb}
\usepackage{rotate}
\usepackage{color}
\usepackage[ps2pdf=true]{hyperref}
\usepackage{bm}

\DeclareFontFamily{OT1}{pzc}{}
\DeclareFontShape{OT1}{pzc}{m}{it}%
            {<-> s * [1.10] pzcmi7t}{}
\DeclareMathAlphabet{\mathscr}{OT1}{pzc}%
                                {m}{it}

\newcommand{\be}{\begin{equation}}
\newcommand{\ee}{\end{equation}}
\newcommand{\bea}{\begin{eqnarray}}
\newcommand{\eea}{\end{eqnarray}}
\def\ba#1\ea{\begin{align}#1\end{align}}
       
\newcommand{\refeq}[1]{Eq.~(\ref{eq:#1})}          
\newcommand{\refeqs}[2]{Eqs.~(\ref{eq:#1})--(\ref{eq:#2})}          
          
\newcommand{\reffig}[1]{Fig.~\ref{fig:#1}}

\newcommand{\vs}{\nonumber\\}       
\newcommand{\refsec}[1]{Sec.~\ref{sec:#1}}          
          
\definecolor{RedWine}{rgb}{0.743,0,0}
\definecolor{RoyalBlue}{rgb}{0.25,.41,.88}
\definecolor{Green}{rgb}{0.1,0.5,0.2}

\renewcommand{\v}[1]{\mathbf{#1}}

\newcommand{\vth}{\bm{\theta}}

\newcommand{\vk}{\v{k}}

\newcommand{\<}{\langle}
\renewcommand{\>}{\rangle}

\renewcommand{\k}{\kappa}

\newcommand{\gobs}{\gamma^{\rm obs}}
\newcommand{\kobs}{\kappa^{\rm obs}}

\newcommand{\vkhat}{\v{\hat{k}}}

\renewcommand{\l}{\ell}

\renewcommand{\d}{\delta}

\newcommand{\g}{\gamma}
\newcommand{\D}{\Delta}

\newcommand{\vl}{\bm{\l}}

\newcommand{\Mpch}{\,{\rm Mpc}/h}

\newcommand{\Om}{\Omega_m}

\newcommand{\s}{\sigma}

\newcommand{\chib}{\bar{\chi}}

\newcommand{\M}{\mathcal{M}}

\newcommand{\fNL}{f_{\rm NL}}

\def\fnl{{f_{\rm{NL}}}}

\newcommand{\mnras}{MNRAS}

\newcommand{\apjl}{Astrophys. J. Lett.}

\def\ie{{\em i.e.}}
\def\eg{{\em e.g.}}

\begin{document}

\title{
Primordial Non-Gaussianity and the Statistics of 
Weak Lensing and other Projected Density Fields
}

\author{Donghui Jeong}
\affiliation{Theoretical Astrophysics, California Institute of
    Technology, Mail Code 350-17, Pasadena, California  91125}
\author{Fabian Schmidt}
\affiliation{Theoretical Astrophysics, California Institute of
    Technology, Mail Code 350-17, Pasadena, California  91125}
\author{Emiliano Sefusatti}
\affiliation{Institut de Physique Th\'eorique\\ CEA, IPhT, 91191 Gif-sur-Yvette c\'edex, France\\ CNRS, URA-2306, 91191 Gif-sur-Yvette c\'edex, France}

\begin{abstract}
Estimators for weak lensing observables such as shear and convergence generally have non-linear corrections, which, in principle, make weak lensing power spectra sensitive to primordial non-Gaussianity.  In this paper, we quantitatively evaluate these contributions for weak lensing auto- and cross-correlation power spectra, and show that they are strongly suppressed by projection effects.  This is a consequence of the central limit theorem, which suppresses departures from Gaussianity when the projection reaches over several correlation lengths of the density field, $L_P\sim 55 \Mpch$.  
Furthermore, the typical scales that contribute to projected
bispectra are generally smaller than those that contribute to 
projected power spectra.
Both of these effects are not specific to lensing, and thus affect the statistics of non-linear tracers (\eg, peaks) of any projected density field.  
Thus, the clustering of biased tracers of the three-dimensional density field is generically more sensitive to non-Gaussianity than observables constructed from projected density fields. 
\end{abstract}
\date{\today}

\maketitle

\section{Introduction}
\label{sec:intro}

In recent years, a renewed interest in the effects of primordial non-Gaussianity on the large-scale structure of the Universe has emerged, prompted by the significant effect on the bias of dark matter halos at large scales measured in N-body simulations \cite{DalalEtal08, DesjacquesSeljakIliev2009, GrossiEtal2009, PillepichEtal2010, WagnerVerde2011, letter}. The observation of this effect in redshift surveys would be able to provide an 
independent confirmation of a possible detection of primordial non-Gaussianity from the anisotropies of the cosmic microwave background (CMB). Such a detection 
would open a completely new perspective on inflation and the 
high-energy physics 
of the early Universe \citep[see, for instance,][]{KomatsuEtal2010}.

In fact, a specific type of non-Gaussian initial conditions, \ie ~the {\em local} model of primordial non-Gaussianity \citep{SalopekBond1990, KomatsuSpergel2001}, induces a strongly 
scale-dependent correction to the linear halo bias.  This
correction has been derived using several approaches,
mostly based on the peak-background split formalism 
\citep{DalalEtal08, SlosarEtal, AfshordiTolley08, 
DesjacquesSeljakIliev2009, GP, fsmk} or on the statistics of 
high-peaks \citep{MLB, MV08} (also, see \citep{McDonald2008} 
for a different perspective). 
On the other hand, one can make the simple assumption of a local 
but {\em nonlinear} 
bias relation between the galaxy distribution and the underlying matter 
field \cite{scoccimarro2000}.  When applied to the local model of non-Gaussianity, this yields the same scale-dependent correction as obtained in the former
two approaches \cite{TKM08,DJS}.  

The effect due to nonlinear local biasing shows an especially close analogy with the case of non-linear weak lensing estimators we consider in this paper. A nonlinear but local galaxy bias relation can be expressed by the Taylor expansion \citep{FryGaztanaga1993}
\be
\d_g(x)=b_1\,\d(x)+\frac12 b_2\,\d^2(x)+\cdots\,,
\ee
where $\d_g(x)$ and $\d(x)$ represent, respectively, 
the galaxy and matter density contrasts and 
$b_1$ and $b_2$ are the linear and quadratic bias parameters, 
which are assumed to be constants for a given galaxy sample. 
This leads to the following 
perturbative expansion for the galaxy two-point correlation function:
\ba
\<\d_g(x_1)\d_g(x_2)\> = \:& b_1^2\<\d(x_1)\d(x_2)\>\nonumber\\
& +\frac12 b_1b_2\left[\<\d(x_1)\d^2(x_2)\>+(x_1\leftrightarrow x_2)\right]\nonumber\\
& +\cdots\,,
\label{eq:dgdg_3d}
\ea 
where the second term on the r.h.s. depends on the matter 3-point correlation function or bispectrum. For Gaussian initial conditions, the matter bispectrum induced by gravitational instability leads to negligible corrections in the two-point correlation on large scales. For non-Gaussian initial conditions of the local type, on the other hand, the correction becomes relevant on large scales.  

Since such behavior arises simply from the {\em nonlinear} relation between the observed distribution $\d_g$ and the matter distribution $\d$, we expect similar effects to show up for any other large-scale structure observables where analogous nonlinearities are present. 

Weak lensing measures the three components $\k$, $\g_1$, $\g_2$ of the tidal tensor of the lensing potential $\psi$,
\ba
\psi_{,ij}=\:&\left( \begin{array}{cc} 
 \k+\g_1 & \g_2\\
 \g_2 & \k-\g_1
\end{array}\right)\,,\label{eq:A}\\
\psi(\vth) \equiv\:& \int_0^{\chi_s} d\chi\: \frac{\chi_s-\chi}{\chi \chi_s}\:
\Phi_{-}(\chi\vth; \chi)\,,
\ea
where derivatives are with respect to angular coordinates on the sky (we assume small angles, zero curvature, and a flat sky limit throughout), $\chi$ denotes comoving distance, while $\chi_s$ is the distance to the source galaxy being lensed, and $\Phi_- = (\Psi-\Phi)/2$ is the lensing potential in conformal Newtonian, or longitudinal, gauge.  

While the two components of the shear $\g_i$ can be measured using
galaxy ellipticities, the convergence $\k$ can be measured either
through its effects on the number density of galaxies (magnification
bias), or on galaxy sizes and fluxes.  
In all of these cases, the estimators are not purely linear.  This
is not surprising since \refeq{A} is only a lowest order
approximation to the lensing effect.  Given that the shear $\g$ is a
(two-dimensional) spin-2 field while $\k$ is a scalar, we can generally
write
\ba
\gobs(\vth) =\:& \g(\vth) + c_\g \k(\vth) \g(\vth) + \cdots \label{eq:gobs}\\
\kobs(\vth) =\:& \k(\vth) + c_{\k1} \k^2(\vth)\label{eq:kobs}\\
& + c_{\k2} [\g_1^2(\vth)+\g_2^2(\vth)] + \cdots,\nonumber
\ea
where $\kobs,\gobs$ are the measured convergence and shear, and the dots indicate third- and higher order terms.  For simplicity, we have assumed that all additive and multiplicative biases have been removed so that to lowest order, $\kobs = \k$ and $\gobs = \g$.

In the case of shear measurements from galaxy surveys, corrections to the leading order come from two sources:  the fact that galaxy shapes estimate the reduced shear $g=\gamma/(1-\kappa)$ rather than the shear $\gamma$ \citep{White2005, DodelsonShapiroWhite2006, BernardeauBonvinVernizzi2010}; and \emph{lensing bias} \cite{sizepaperII}, the fact that we preferentially select lensed galaxies in a flux- and size-limited source galaxy sample.  Following the estimate of \cite{sizepaperI}, these two effects
add up, yielding $c_\g$ roughly between $2$ and $3$.

Examples of measurements of the convergence include using the number density of background galaxies via the magnification and size bias effects \citep{Broadhurst1994, ScrantonEtal2005}, and using the sizes and  other measured characteristics of galaxies 
\citep{jain02, BertinLombardi, RozoSchmidt}. All these estimators in fact measure the magnification $\mu$,
\be\label{eq:mu}
\mu(\vth)=\frac{1}{(1-\k)^2-|\g|^2} = 1 + 2\k + 3\k^2 + |\g|^2 + \cdots
\ee
Hence, in the ideal case we expect $c_{\k1} = 3/2$ and $c_{\k2}=1/2$. These values will likely be modified in practice due to galaxy selection effects similar to those present in the shear. Note that it is often possible to vary the non-linearity coefficients $c_\g,c_\k$ experimentally, for example by applying different cuts on the source galaxy sample.  

In analogy with \refeq{dgdg_3d}, the two-point correlation of a non-linear estimator such as \refeqs{gobs}{kobs} receives corrections from three- and higher point functions of the underlying density field.  This is easy to see \eg ~for the shear two-point function (neglecting the spinor indices for simplicity):
\ba
\<\gobs(\vth)\gobs(\vth')\> =\:& \<\g(\vth)\g(\vth')\> \label{eq:shearcorr}\nonumber\\
& + 2 c_\g \<\k(\vth)\g(\vth)\g(\vth')\> + \cdots
\ea
The leading correction term is a three-point function of shear and convergence,
evaluated in the limit where two of the three vertices coincide.  
In \cite{Dodelson:2005ir,Shapiro2009, sizepaperII}, this contribution was 
investigated for non-Gaussianities from the non-linear evolution of the matter density.  
However, primordial non-Gaussianity also modifies the shear and convergence power spectra
in the same way. Throughout this paper, we shall focus on the local model of 
non-Gaussianity, where the bispectrum of primordial curvature perturbations, 
evaluated at last scattering, is given by
\be
B_\phi(k_1,k_2,k_3) = 2\fnl [ P_\phi(k_1)P_\phi(k_2) + {\rm perm.} ].
\vspace*{0.2cm}
\label{eq:Bloc}
\ee
Our conclusions however are broadly valid for any primordial bispectrum shape. 

It will be useful to compare the impact of non-Gaussianity on weak lensing 
correlations with the analogous effects on the clustering of large-scale
structure tracers such as galaxies or halos.  As shown in \cite{DalalEtal08}, 
tracers with a linear (Eulerian) bias $b_1$ acquire a scale-dependent 
bias correction in the presence of local non-Gaussianity given by
\be
\label{eq:deltab_localfnl}
\Delta b_1(k;z) =
\frac{2\fnl(b_1-1)\delta_c}{\M(k,z)},
\ee
where $\d_c = 1.686$ is the critical density for the spherical collapse
in the flat matter dominated universe. The function 
$\M(k,z)$, which relates the matter density fluctuations to the initial 
curvature perturbations as $\d(\vk)=\M(k,z)\phi(\vk)$, is given by
\be
\M(k,z) \equiv  \frac{k^2\: T(k)}{C_P} D(z).
\ee
where $T(k)$ is the matter transfer function, $C_P \equiv 3/2\:\Om H_0^2$, 
and $D(z)$ is the linear growth function normalized to the scale factor at last 
scattering (at which in our 
convention $\fNL$ is defined), that is $D(z_{ls})=1/(1+z_{ls})$.

Throughout this paper, we shall also assume the Limber approximation and work in the small angle limit, where the spherical harmonic transform is reduced to the two-dimensional Fourier transform.  This very useful approximation does not significantly influence our results \cite{jeong/komatsu/jain:2009}. We also assume that sources reside at a fixed redshift.  Our fiducial cosmology is defined through the maximum likelihood cosmological parameters in Table 1 (``WMAP5+BAO+SN'') of \citet{komatsu/etal:2009}.  

We begin in \refsec{shear} by studying the impact of non-Gaussianity on shear and convergence power spectra.  We then investigate the cross-correlation of 
shear and convergence with large-scale structure (LSS) tracers in \refsec{gglensing}. \refsec{peaks} extends the arguments to the general case of the clustering of peaks (or more generally nonlinear tracers) identified in projected density fields.  We conclude in \refsec{disc}

\section{Shear and convergence power spectra}
\label{sec:shear}

We first consider the impact of primordial non-Gaussianity on large-scale shear and convergence power spectra.  By Fourier transforming \refeq{shearcorr} and the analogous equation for the convergence, we obtain the observed shear and convergence power spectra:
\begin{widetext}
\ba
C_{\gobs}(\l) =\:& C_\k(\l) 
+ 2 c_\g \!\int\!\frac{d^2\l_1}{(2\pi)^2}
\cos 2\phi_{\l_1} B_\k(\l_1,|\vl-\vl_1|,\l)\label{eq:dCg}\,,\\
C_{\kobs}(\l) =\:& C_\k(\l) + 2 \!\int\!\frac{d^2\l_1}{(2\pi)^2}
\left[ c_{\k1} + 2 c_{\k2} \cos 2(\phi_{\l_1}-\phi_{\l-\l_1}) \right ] 
B_\k(\l_1,|\vl-\vl_1|,\l)\label{eq:dCk}\,.
\ea
\end{widetext}
Here, $B_\k$ is the convergence bispectrum, and we have aligned $\vl_1$ so that $\phi_\l=0$.  Further, we have used that in Fourier space $\g(\vl) = \g_1(\vl) + i \g_2(\vl) = \exp(2i\phi_\l) \k(\vl)$, and hence the leading order shear power spectrum is equal to the leading order convergence power spectrum $C_\k(\l)$. The convergence power spectrum and bispectrum are related to the corresponding matter correlators $P_m,\;B_m$ via 
\bea
C_\k(\l)\! &=&\! C_P^2 \!\int_0^{\chi_s}
\frac{d\chi}{\chi} \frac{W_L^2(\chi_s,\chi)}{\chi \: a^2(\chi)}  P_m(\l/\chi;\chi)\,,\  \  \  \  \\
B_\k(\l_1,\l_2,\l_3)\! &=& \!C_P^3\! \int_0^{\chi_s}
\frac{d\chi}{\chi} \frac{W_L^3(\chi_s,\chi)}{\chi^3\:a^3(\chi)}
\vs & & \qquad\times 
B_m \!\left (\frac{\l_1}{\chi},\frac{\l_2}{\chi},\frac{\l_3}{\chi}; \chi \right ),
\eea
where $W_L(\chi_s,\chi)\equiv \chi/\chi_s (\chi_s-\chi)$ denotes the lensing kernel. 

In Eulerian Perturbation Theory (PT), the leading order expression (``tree-level'') of the matter bispectrum, valid at large scales, is given by the sum of a primordial component, $B_{I}$, present for non-Gaussian initial conditions, and a contribution $B_G$ due to second-order corrections to the matter density induced by 
gravitational instability.  We have
\be
B_m = B_I+ B_G+...\,,
\ee
where the dots represent higher-order corrections in PT.  The primordial component $B_I$ is directly related to the primordial curvature bispectrum 
$B_\phi$ via
\ba\label{eq:Bphi_to_Bm}
B_I(k_1, k_2, k_3; z) =\:& \M(k_1,z)\M(k_2,z)\M(k_3,z)\vs
&\times B_\phi(k_1,k_2,k_3)\,,
\ea
while the non-linear component $B_G$ is given by
\ba
B_G(k_1, k_2, k_3; z) =\:& 2 F_2(\vk_1,\vk_2)P_L(k_1;z)P_L(k_2;z)\vs
& +(2~{\rm perm.})\,,
\ea
where $P_L(k;z) = \M(k,z)^2 P_\phi(k)$ is the linear matter power spectrum,
and
\ba
F_2(\vk_1,\vk_2) =\:& 5/7+(1/2)\vkhat_1\cdot\vkhat_2(k_1/k_2+k_2/k_1)\vs
& +(2/7)(\vkhat_1\cdot\vkhat_2)^2.
\ea
Note that at small scales, additional perturbative corrections become relevant and, in general, 
it is not possible to separate out the purely primordial components in the matter bispectrum \citep{Sefusatti2009, SefusattiCrocceDesjacques2010}. We shall return to this issue below.

From the non-Gaussianity viewpoint, \refeqs{dCg}{dCk} come as no surprise:  neglecting the distinction between $\g$ and $\k$ for the moment, \refeqs{gobs}{kobs} say that the observed shear and convergence are  \emph{biased} estimates of the true $\g$, $\k$ with linear bias $b_1=1$ and quadratic bias parameters $b_2 = c_\g, c_\k$.  Hence, \refeqs{dCg}{dCk} are analogous to the expression for the non-Gaussian halo power spectrum \cite{TKM08}, with two differences:  first, we are dealing with the projected density field $\kappa$;  second, the relation between shear and convergence leads to cosine factors in the integral.  Note that on large scales, the bispectrum $B_\k$ is evaluated in the squeezed limit, since typically $\l \ll \l_1$.  This is again similar to the halo clustering case.

\begin{figure}[t!]
\centering
\includegraphics[width=0.48\textwidth]{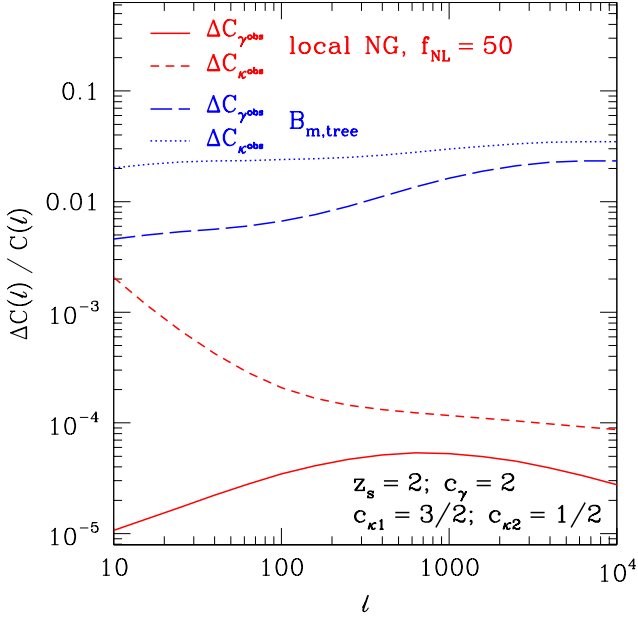}
\caption{
Relative correction to the shear and convergence power spectra from primordial NG of the local type (red solid/dashed), and the tree-level matter bispectrum from non-linear evolution (blue long-dashed/dotted).  Here, we have assumed sources located at $z_s=2$ and non-linear parameters as indicated in the figure (see \refsec{intro}).  
\label{fig:dCk}}
\end{figure}

\reffig{dCk} shows the relative amplitude of the correction to the shear and convergence power spectra from a numerical evaluation of \refeqs{dCg}{dCk} when using the primordial bispectrum of the local type.  We also show the tree-level bispectrum from non-linear evolution which contributes at the percent-level to $C_{\gobs}$ and $C_{\kobs}$.   Clearly, the contributions from primordial non-Gaussianity are strongly suppressed:  they are always below $10^{-4}$ for the shear, and only reach $10^{-3}$ at the very largest scales for the convergence.  This is in stark contrast to the results for the halo power spectrum \cite{DalalEtal08,SlosarEtal} where order unity corrections are observed 
for this type of non-Gaussianity on large scales.  

In order to understand where this suppression comes from, we make an order of magnitude estimate of \refeqs{dCg}{dCk}. We approximate the convergence power and bispectrum as
\be
C_\k(\l) \sim C_P^2 \,\chib\, x_P^2\, P_L(\l/\chib;\chib)\,,
\label{eq:Ckap}
\ee
and 
\be
B_\k(\l_1,\l_2,\l_3) \sim C_P^3\, x_B^3 \,
B_I \left (\frac{\l_1}{\chib},\frac{\l_2}{\chib},\frac{\l_3}{\chib}; \chib \right )\,.
\ee
Here, $\chib$ is an effective lens distance, and $x_P$, $x_B$ are dimensionless
geometrical factors of order unity which we leave unspecified for the moment.  
On large scales, the bispectrum is evaluated in the squeezed limit, 
$k_1 \gg k$.  In the squeezed limit, \refeq{Bloc} asymptotes approximately to
\ba
B_I(k_1, k, |\vk_1\!-\!\vk|) \:&= 2\,\fNL\, \M(k) \M(k_1) \M(|\vk_1\!-\!\vk|)\nonumber\\
 &\quad \times P_\phi(k) \left [ P_\phi(k_1)\! +\! P_\phi(|\vk_1\!-\!\vk|)\right ]\  \  
 \label{eq:Bsqueezed}\\
\:& \sim 4\,\fNL\, \M(k) P_\phi(k) P_L(k_1)\label{eq:Bsq2}
\ea
neglecting the third permutation which is suppressed.  In the second 
approximate equality, we also set $|\vk_1-\vk| \approx k_1$.  Further,
for the moment all power spectra are assumed to be evaluated at $z(\chib)$.  
After some manipulation, we arrive at the following estimates:
\ba
\D C_{\gobs}(\l) \approx\:& 0\,,\\
\D C_{\kobs}(\l) \approx\:& 8 \fNL \frac{(c_{\k1}+c_{\k2}) \sigma_\k^2}{C_P\chib^2 x_B} \left(\frac{x_B}{x_P}\right)^2\vs
 &\times \M^{-1}(\l/\chib)  C_\k(\l)\,,\label{eq:dCsim1}
\ea
where
\be
\sigma_\k^2 = \int \frac{d^2\l}{(2\pi)^2} C_\k(\l).\label{eq:sk}
\ee
The $\cos 2\phi_{\l_1}$ factor in \refeq{dCg} leads to a complete cancelation
of the effect on $C_{\gobs}$ in this approximation, while the 
phase factor in the second term of \refeq{dCk} becomes unity.  
Further, note that in reality there is a high$-\l$ cut-off in \refeq{sk} due to
the resolution or pixel size of the shear survey.  As long as this cut-off
is less than $\sim$ arcmin-scale, however, the quantitative results do not
depend sensitively on the resolution.  

\begin{figure}[t!]
\centering
\includegraphics[width=0.48\textwidth]{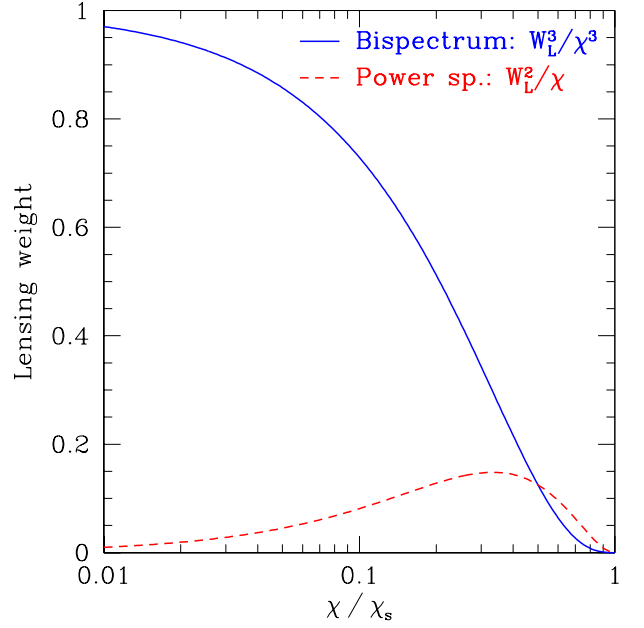}
\caption{Effective lensing weight functions for the shear power spectrum
(red dashed), and the shear bispectrum (blue solid).  
\label{fig:W}}
\end{figure}

Using \refeq{Ckap}, \refeq{dCsim1} can be further simplified to yield
\be
\frac{\D C_{\kobs}(\l)}{C_\k(\l)} \approx 8  \fNL \frac{c_\k}{\l^2 T(\l/\chib)} \frac{\sigma_\k^2}{x_B}
\left(\frac{x_B}{x_P}\right)^4,
\label{eq:dCsim2}
\ee
where $c_\k = c_{\k1}+c_{\k2}$.  
Note the appearance of an $\l^{-2}$, just as a factor of $k^{-2}$ appears
in the halo biasing in the local model of NG \cite{DalalEtal08}.  
In fact, it is instructive to compare \refeq{dCsim2} to a similar estimate
for the angular power spectrum of some biased tracer ``$h$''.  Using the Limber 
approximation, we have
\be
C_{hh}(\l) = \int\frac{d\chi}{\chi} \frac{F_h(\chi)^2}{\chi} P_{hh}\left(\frac{\l}{\chi}; \chi\right),
\ee
where $F_h(\chi)$ is the selection function, normalized to unity in comoving
distance.  Let us now assume that the tracer is localized in a narrow redshift
slice around a comoving distance $\chib$.  Given that 
$P_{hh}(k) = [b_1 + \D b(k)]^2 P_m(k)$ where $\Delta b$ is given by
\refeq{deltab_localfnl} in the local model of NG, we can approximately
write the leading correction to $C_{hh}$ as
\be
\frac{\D C_{hh}(\l)}{C_{hh}(\l)} \approx 2 \fnl \frac{b_1-1}{b_1} \d_c
\frac{C_P \chib^2}{D(\bar z)\,\l^2 T(\l/\chib)},
\label{eq:dChsim}
\ee
where $\bar z = z(\chib)$. 
Assuming that $\chib$ is a cosmological distance, all factors in \refeq{dChsim}
are in fact of order unity.  Comparing \refeq{dChsim} with \refeq{dCsim2}
shows that the suppression of the impact of non-Gaussianity
on weak lensing power spectra comes from two sources:  first, it
is suppressed by $\s_\k^2 \sim 10^{-3}$.  As we will show in \refsec{peaks},
this suppression factor is due to the projection of the density field over
many correlation lengths, and thus ultimately a consequence of the central
limit theorem.   Second, the projection kernel for the bispectrum strongly
prefers small lens distances, so that $x_B \ll x_P \sim 0.5$.  To
illustrate this, \reffig{W} shows the redshift
weighting function of the shear bispectrum and power spectrum in comparison.  
Clearly, most of the contribution to the shear bispectrum comes from
low redshifts, $\chi/\chi_s \lesssim 0.2$.  In other words, for a survey with $z_s=2$
($\chi_s \sim 3700\Mpch$), the shear bispectrum hardly probes scales above
$700\Mpch$.  Since the effect (at least from local NG) peaks on large scales,
this is a severe limitation.  

\begin{figure}[t!]
\centering
\includegraphics[width=0.48\textwidth]{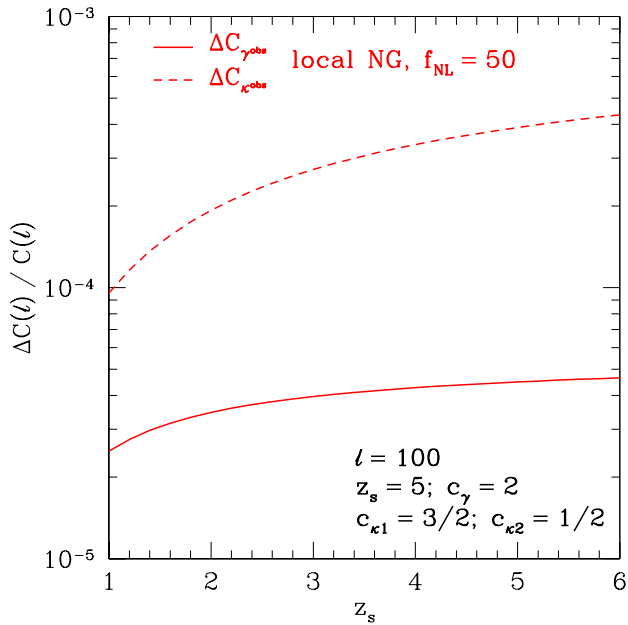}
\caption{Relative amplitude of the non-linear lensing correction from
primordial non-Gaussianity to the shear and convergence power spectra as 
function of the source redshift $z_s$, for a fixed multipole $\l=100$. 
The other parameters are at the same values as for \reffig{dCk}.  
\label{fig:dC-vs-z}}
\end{figure}

Finally, choosing some 
numbers, $\s_\k^2 = 10^{-3}$, $x_B=0.2$, $x_P=0.5$, and
setting $T=1$, we have
\be
\frac{\D C_{\kobs}(\l)}{C_\k(\l)} \sim 10^{-4} \left(\frac{\fNL}{100}\right) 
\left(\frac{\l}{100}\right)^{-2}.
\label{eq:oom}
\ee
Comparing with \reffig{dCk}, we see that the order-of-magnitude estimate 
predicts the right amplitude to within a factor of a few.  We also see
the $\l^{-2}$ behavior for $\D C_{\kobs}$ on large scales, and that
$\D C_{\gobs}$ is indeed strongly suppressed for small $\l$ due to the cosine 
factor in \refeq{dCg}.  
The restriction to small scales due to the lensing projection can be
somewhat mitigated by going to larger source redshifts.  \reffig{dC-vs-z}
shows the evolution of the non-linear corrections with source redshift.  
While the corrections, especially $\D C_{\kobs}$, increase with $z_s$,
the corrections remain much smaller than $10^{-3}$ at $\l=100$.  

\subsection{Impact of non-linearities on small scales}

So far, we have only considered the leading order (tree-level) contribution to the matter bispectrum from primordial non-Gaussianity.  A simple estimate of the effect of non-linearities can be obtained directly from \refeq{dCsim2}, by replacing $\sigma_\k^2$ with the non-linear variance of the convergence. Since $\s_{\k,\rm NL}^2 \sim 5 \s_\k^2$ (estimated using the non-linear matter power spectrum from \texttt{halofit} \cite{halofit}), non-linearities are expected to increase $\D C_\kobs(\l)$ by the same factor.  
We can obtain a somewhat more sophisticated estimate by using the fact that, in the local model of non-Gaussianity, a long wavelength (linear) perturbation $\phi_l$ acts to increase the variance of small-wavelength matter perturbations $\d_{L,s}$ \cite{DalalEtal08, SlosarEtal}:
\be
\<\d_{L,s}^2\> \rightarrow \<\d_{L,s}^2\> \: (1 + 4\fNL \phi_l)\,,
\ee
where all perturbations here are evaluated at early times, {\em i.e.} in Lagrangian space (indicated by subscript ``$L$''). This is clearly reflected in the approximate 
squeezed-limit expression for $B_m$, \refeq{Bsq2} (since $\M(k) P_\phi(k) = \<\d_l\phi_l\>$).  

\begin{figure}[t!]
\centering
\includegraphics[width=0.48\textwidth]{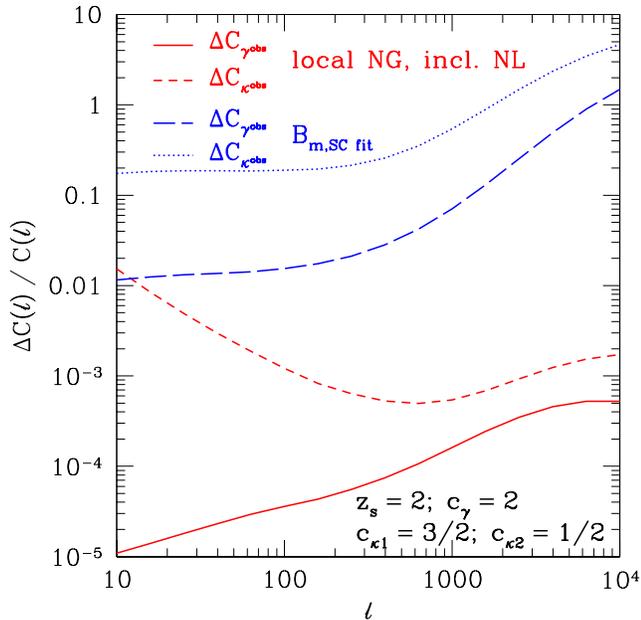}
\caption{
Same as \reffig{dCk}, but with non-linear corrections included. For the corrections from primordial non-Gaussianity (red solid/dashed), we use the matter bispectrum from \refeq{B_NL} with $P_{\mathrm NL}(k)$ given by \texttt{halofit} \cite{halofit}.  For the corrections from non-linear evolution (blue long-dashed/dotted), we use the fitting formula from \cite{SC}.  All other parameters as in \reffig{dCk}. 
\label{fig:dCkNL}}
\end{figure}

In other words, a given region in such a non-Gaussian Universe with a fixed value of $\fNL\phi_l > 0$ is statistically equivalent to a region in a Gaussian Universe with a slightly higher amplitude of the primordial power spectrum.  Thus, we can model the change in the statistics of the late-time, non-linear small-scale modes (in Eulerian space) in the presence of a fixed large-scale mode as
\ba
\<\d_{E,s}(k)\d_{E,s}(-k)\> \rightarrow\:& \<\d_{E,s}(k)\d_{E,s}(-k)\> \nonumber\\
&\times \left ( 1 +  4\fNL \phi_l \frac{\partial P_{\rm NL}(k)}{\partial \ln\mathcal{A}_s} \right)\,,
\ea
where $P_{\rm NL}(k)$ is the non-linear matter power spectrum in a $\Lambda$CDM
cosmology with Gaussian initial conditions.  
Specifically, we estimate the contribution to
the non-linear matter bispectrum due to primordial non-Gaussianity in
the squeezed limit ($k_1\gg k$) as
\ba
B_{I,\rm NL}(k_1, k, k') \:&\simeq 2\fNL \M(k) \M(k_1) \M(k')\vs
\:&\quad\times P_\phi(k) \left [ \M^{-2}(k_1) \frac{\partial P_{\rm NL}(k)}{\partial \ln\mathcal{A}_s}\right.\nonumber\\
\:&\qquad\qquad \left. +   
\M^{-2}(k') \frac{\partial P_{\rm NL}(k')}{\partial \ln\mathcal{A}_s}
\right ]\,, \label{eq:B_NL}
\ea
with $\vk'\equiv \vk_1-\vk$.  
Since $P_L \propto \mathcal{A}_s$, this equation recovers \refeq{Bsqueezed}
when all $k$-vectors are in the linear regime.  We use the \texttt{halofit}
prescription to evaluate the derivatives of $P_{\rm NL}$.  
\reffig{dCkNL} shows the
correction to the shear and convergence power spectra when using the
matter bispectrum from \refeq{B_NL}.  Note that for $\l \gtrsim 400$ 
the squeezed limit is not a good assumption anymore and \refeq{B_NL}
loses its validity.   We see that $\D C_{\kobs}$ is indeed
boosted by matter non-linearities by a factor of $\sim 8$ on large scales, comparable to the
simpler estimate using \refeq{dCsim1}.  On the other hand, $\D C_{\gobs}$
is still suppressed on large scales and remains insignificant.  
\reffig{dCkNL} also shows the correction purely from non-linear evolution
for Gaussian initial conditions, calculated using the bispectrum fitting
formula from \cite{SC} combined with \texttt{halofit} (this is the same
prescription as adopted in \cite{Dodelson:2005ir,Shapiro2009, sizepaperII}).  
This contribution
is clearly boosted as well by including non-linearities on small scales, 
so that it still dominates over the correction from primordial
non-Gaussianity up to very large scales.

\section{Galaxy-galaxy lensing}
\label{sec:gglensing}

We now consider the cross-correlation between a large scale structure tracer $h$ (such as galaxies or galaxy clusters) and weak lensing shear $\g$ and convergence $\k$. 
Such cross correlations, called \textit{galaxy-galaxy} lensing, probe the tracer-mass cross power spectrum. As derived in \cite{jeong/komatsu/jain:2009,NamikawaEtal}, the two-point angular power spectrum between tracer and lensing 
convergence is to leading order in the lensing quantities given by
\begin{align}
\nonumber
C_{h\kappa}(\l)
=\:&
C_P
\left[
b_1(z_L)+\Delta b\left(k=\l/\chi_L;z_L\right)
\right]
\\
& \times (1+z_L)\frac{W_L(\chi_s,\chi_L)}{\chi_L^2}P_m\left(\l/\chi_L;z_L
\right),
\label{eq:Clgk0}
\end{align}
Here, we have again used the Limber approximation.  $C_{h\k}$ can be estimated either through convergence estimators or through shear, by using the relation
between $\k$ and $\g$ in Fourier space.  

The derivation of the nonlinear lensing corrections proceeds in close analogy to \refsec{shear}, and we obtain:
\ba
\D C_{h\gobs}(\l) =\:& c_\g
\int\frac{d^2\l_1}{(2\pi)^2} \cos 2\phi_{\l_1}\nonumber\\
\:&\quad\quad\times B_{h\k\k}(\l,\l_1,|\vl-\vl_1|)\,,\label{eq:dChg}\\
\D C_{h\kobs}(\l) =\:& c_{\k1} \int\frac{d^2\l_1}{(2\pi)^2} B_{h\k\k}(\l,\l_1,|\vl-\vl_1|)\vs
\:& + c_{\k2} \int \frac{d^2\l_1}{(2\pi)^2}
\cos 2(\phi_{\l_1}-\phi_{\l-\l_1})\nonumber\\
\:&\quad\quad\quad\quad\times B_{h\k\k}(\l_1,|\vl-\vl_1|,\l)\,.\label{eq:dChk}
\ea
Here, the halo-$\k$-$\k$ bispectrum is defined through
\be\hspace{-0.1cm}
\left\langle
h(\bm{\l}_1)
\k(\bm{\l}_2)
\k(\bm{\l}_3)
\right\rangle
\equiv
(2\pi)^2 \delta_D(\bm{\l}_{123})B_{h\k\k}(\l_1,\l_2,\l_3)\,.
\ee
In the Limber approximation (and assuming linear biasing), it is given in 
terms of the matter bispectrum by:
\ba
B_{h\k\k}(\l_1,\l_2,\l_3;z_L) = \:& 
\left[C_P W_L(\chi_s,\chi_L)(1+z_L)\right]^2
\frac{1}{\chi_L^{4}}\vs
\:& \times [b_1+\D b(\l_1/\chi_L;z_L)]\vs
\:& \times B_m\left(\frac{\l_1}{\chi_L},\frac{\l_2}{\chi_L},\frac{\l_3}{\chi_L};z_L\right).
\ea
Again, we assume sufficiently large scales so that $B_m = B_I + B_G$.  
Note that we have two contributions of order $\fNL$, $\propto b_1 B_I$ and
$\propto \D b_1 B_G$, and a contribution of order $\fNL^2$, $\propto \D b_1 B_I$.  

\begin{figure}[t!]
\centering
\includegraphics[width=0.48\textwidth]{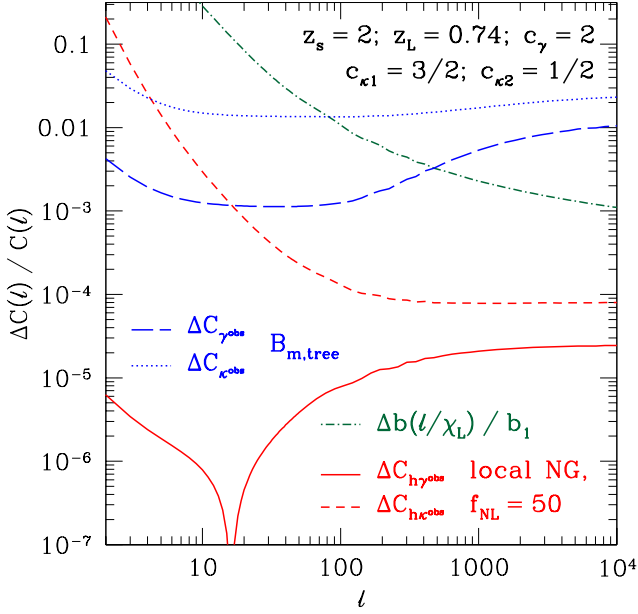}
\caption{Relative correction to the halo-shear and halo-convergence cross 
power spectra from primordial NG of the local type (red solid/dashed).    
The blue long-dashed/dotted lines show the correction from the tree-level matter bispectrum from non-linear evolution in the absence of primordial non-Gaussianity. Here, we have assumed a lens redshift of $z_L=0.74$ and sources at $z_s=2$ so that $\chi_s=2\chi_L$.  
The dash-dotted line shows the correction to $C_{h\g}, C_{h\k}$ from the scale-dependent halo bias \refeq{deltab_localfnl} [via \refeq{Clgk0}],
assuming $b_1=2$. }
\label{fig:dCh}
\end{figure}

\reffig{dCh} shows the numerical result for \refeqs{dChg}{dChk}, assuming lens galaxies at $z_L=0.74$ and source galaxies at $z_S=2.0$. The redshift of the source galaxies is chosen by requiring $\chi_s = 2 \chi_L$. We assume that the linear bias parameter of lens galaxies is $b_1=2$, and the amplitude of non-Gaussianity is given by $\fnl=50$.   We adopt the same values for the non-linearity coefficients $c_\g, c_{\k1}, c_{\k2}$ as in the last section.  
We again see that the correction to $C_{h\g}$ is suppressed, while $C_{h\k}$ receives a correction that strongly increases towards large scales;  in fact, at $\l \lesssim 10$, the term $\propto \D b\: B_I \propto \l^{-4}$ becomes larger than the contribution from the tree level matter bispectrum for these parameter values. Following similar steps to \refsec{shear}, we can obtain an order of magnitude estimate of the nonlinear correction to $C_{h\k}$ from the local primordial bispectrum $B_I$ (for $C_{h\g}$ we again have a cosine factor which leads to a 
cancelation in the large-scale limit).  
Assuming the approximation of \refeq{Bsqueezed} we can write $\D C_{h\kobs}$ as
\ba
\Delta C_{h\kobs}(\l)
\approx\:& \frac{4\, c_\k\, b_1\, \fnl}{\M\left(\l/\chi_L;\:z_L\right)} C_P^2\, \left[\frac{D(z_L)}{D(0)}\right]^2 \,P_L\left(\frac{\l}{\chi_L};z_L\right)\vs
&\times  W^2_L(\chi_s,\chi_L) \frac{(1+z_L)^2}{\chi_L^2} L_P,
\ea
where $c_\k$ is defined below \refeq{dCsim2}.  Hence the relative correction
to the tracer-convergence cross-correlation becomes
\ba
\frac{\Delta C_{h\kappa}(\l)}{C_{h\kappa}(\l)}
\approx\:& 4\, c_\k\, \fnl\, (1+z_L)\, D(z_L)\vs
\:&\times \frac{C_P^2\, \chi_L^2\, W_L(\chi_s,\chi_L)\, L_p}
{\l^{\,2}\, T(\l/\chi_L)\, D^2(0)}\,.
\label{eq:ratio0}
\ea
Note that the bias factor $b_1 + \D b(l/\chi_L)$ drops out.  
In these expressions, a new length scale appears due to the projection,
\be
L_p \equiv \int \frac{d^2 k}{(2\pi)^2} P_m(k,z=0)\,.
\ee
$L_p$ can be seen as the one-dimensional coherence length of the density field at $z=0$, in the sense that the variance of the density field projected along a thin slab of thickness $\D\chi$ at redshift $z$ is given by
\be
\s_{\d\:\rm proj}^2 = \left[\frac{D(z)}{D(0)}\right]^2 \frac{L_p}{\D\chi}.
\label{eq:sigproj}
\ee
In our fiducial cosmology, $L_p = 54.6\:\mathrm{Mpc}/h$.

\begin{figure}[t!]
\centering
\includegraphics[width=0.48\textwidth]{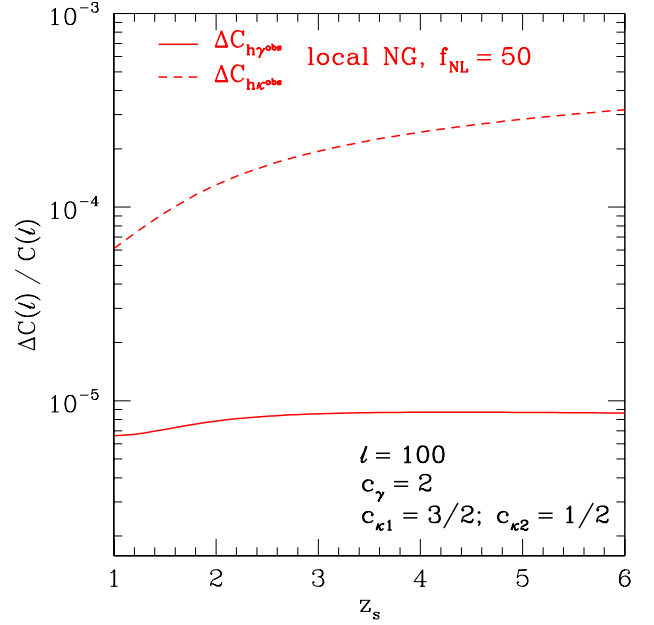}
\caption{Relative amplitude of the non-linear lensing correction 
from primordial non-Gaussianity to the halo-shear and halo-convergence
cross-correlation, as function of the lens redshift $z_L$,
for a fixed multipole $\l=100$.  The other parameters are the same as for
\reffig{dCh}.
\label{fig:dCh-vs-z}}
\end{figure}

Since on large scales $T(\l/\chi_L) \approx 1$, \refeq{ratio0} again 
recovers the 
$\l^{-2}$ scaling of the non-linear correction, similar to what was
found for the weak lensing power spectrum.  Compared to the change in
$C_{h\k}(\l)$ due to the scale-dependent halo bias in \refeq{Clgk0}
(green dot-dashed line in \reffig{dCh}), 
which is exactly $1/2$ the effect on the halo angular power spectrum 
\refeq{dChsim}, the effect of the non-linear lensing correction is 
suppressed by a factor of
\be
C_P W_L(\chi_L,\chi_s) L_p \sim L_p H_0 \sim 0.02,
\ee
where we have assumed that $W_L$ is of order the horizon scale.  While this is a significant reduction, the projection effect does not suppress the correction in galaxy-galaxy lensing quite as severely as it does for the shear power spectrum.  This is mainly because in the cross-correlation with biased tracers the dominant contribution arises at $z_L$, rather than at low redshifts as in the lensing auto-correlation. Comparing with \reffig{dCh}, the order of magnitude prediction again matches quite well.  

\reffig{dCh-vs-z} shows the amplitude of the corrections to $\D C_{h\k}$,
$\D C_{h\g}$ as a function of source redshift.  Here, we have chosen
the lens redshift such that $\chi_s = 2\chi_L$ in each case; thus
maximizing the lensing kernel $W_L$.  The scaling in redshift
is very similar to that seen in the weak lensing power spectra,
\reffig{dC-vs-z}.

\section{General statements on Peak clustering in Projected Fields}
\label{sec:peaks}

The results derived in the last two sections allow us to make some interesting and general statements on the effectiveness of using the clustering of peaks (or, more generally, non-linear tracers) identified in two-dimensional, projected density fields.  Applications of this include shear peaks as well
as peaks identified in diffuse background maps.  

Consider a general projected density field
\be
\lambda(\vth) \equiv \int d\chi F_\lambda(\chi) \d(\chi\vth;\chi)\,,
\label{eq:lambda}
\ee
where $F_\lambda(\chi)$ is a filter function normalized to unity in $\chi$. Using the Limber approximation, the angular power spectrum of $\lambda$ is then straightforwardly obtained as 
\be
C_\lambda(\l) = \int\frac{d\chi}{\chi} \frac{F_\lambda^2(\chi)}{\chi} P_m\left(\frac{\l}{\chi};\chi\right).
\ee
Now assume we identify peaks in $\lambda$; that is, we apply some nonlinear transformation so that the peak density is perturbatively given by
\be
\lambda_{\rm pk}(\vth) = c_1 \lambda(\vth) + \frac{c_2}{2} \lambda^2(\vth) + \cdots
\ee
The simplest example is thresholding, {\em i.e.} $\lambda_{\rm pk}(\vth) = \Theta(\lambda(\vth)-\lambda_c)$, where $\Theta$ is the Heaviside function.  If $\lambda_c \gg \s_\lambda$, $\s_\lambda$ being the r.m.s. fluctuation of $\lambda$ given by
\be
\s_{\lambda}^2 = \int \frac{d\chi}{\chi} \frac{F_\lambda^2(\chi)}{\chi}
\int \frac{d^2\l}{(2\pi)^2} P_m(\l/\chi;\chi)\,,
\ee
$c_1,\:c_2$ attain the well-known values \cite{MLB}
\be
c_1 = \frac{\lambda_c}{\s_\lambda^2};\quad c_2 = c_1^2\,.
\label{eq:cl}
\ee
The following considerations are completely independent of the precise
peak definition and value of $c_2$, however.  Analogous to our derivations
in \refsec{shear} and~\ref{sec:gglensing}, the angular power spectrum
of peaks can be written as 
\ba
C_{\lambda_{\rm pk}}(\l) =\:& c_1^2\: C_{\lambda}(\l) + \D C_{\lambda_{\rm pk}}(\l)\,,\vs
\D C_{\lambda_{\rm pk}}(\l) =\:& c_1 c_2 \int \frac{d^2\l}{(2\pi)^2} B_\lambda(\l,\l_1,|\l_1-\l|)\,.
\ea
The second term can in principle be used to probe primordial non-Gaussianity.  
We can now use the Limber approximation, and in the large scale
limit make the same set of approximations as in the previous sections. 
As a further simplification, we will assume that the kernel $F_\lambda(\chi)$ is
peaked at some distance $\chib$.  We then obtain the following estimate for the
impact of non-Gaussianity on the clustering of peaks in the projected field $\lambda$:
\ba
\frac{\D C_{\lambda_{\rm pk}}(\l)}{C_{\lambda_{\rm pk}}(\l)} \approx\:& 8\fNL 
\frac{c_2 \s_\lambda^2}{c_1} \frac{C_P \chib^2}{\l^2 T(\l/\chib)} 
\frac{1}{D(\bar z)}\vs
\approx\:& 8\fNL 
\frac{c_2}{c_1} \frac{C_P \chib^2}{\l^2 T(\l/\chib)\,D^2(0)} D(\bar z)
\frac{L_p}{\D\chi_\lambda}\,.
\label{eq:dClsim}
\ea
Here, $\bar z = z(\chib)$ and $\D\chi_\lambda \equiv [\int d\chi F_\lambda^2(\chi)]^{-1}$ is the 
effective width of the projection kernel.  In the second line,
we have introduced the correlation length $L_P$ via \refeq{sigproj}.  

This result summarizes the different expressions for non-Gaussian corrections
to angular power spectra derived in this paper:  the first line of 
\refeq{dClsim} clearly shows a strong similarity to the expression for
halo clustering, \refeq{dChsim}, upon identifying
$c_2\s_\lambda^2 = c_1 \delta_c$ from \refeq{cl}.  Inserting the appropriate
value for the width $\D\chi_L \equiv [C_P(1+z_L)W_L]^{-1}$ of the lensing kernel 
(at fixed source redshift), setting $c_1=1,\;c_2=c_\k$, and dividing by a 
factor of 2 we also recover the expression for galaxy-galaxy lensing
\refeq{ratio0}  (the factor of 2 comes in since \refeq{dClsim} is for the 
auto-correlation while \refeq{ratio0} is for a cross-correlation).  
Finally, we  see that the quantity $\s_\k^2 \sim L_P/\D\chi_L$ is suppressed by
the same projection effect.  

In summary, the clustering of
peaks identified in the \emph{projected} density field is suppressed relative to the
angular clustering of peaks identified in the \emph{three-dimensional} density
field [\refeq{dChsim}] by a factor of $D^2(\bar z)\,L_p / \D\chi_\lambda$.  
In fact, if the kernel is broad,
then the contributions to the bispectrum $B_\lambda$ are dominated by
low redshifts, as we have seen in \refsec{shear}, which further suppresses
$\D C_{\lambda_{\rm pk}}$ beyond \refeq{dClsim}.  Note that the Limber 
approximation and hence \refeq{dClsim} 
break down if $\D\chi_\lambda \lesssim L_p$, {\em i.e.} for very narrow kernels.  

Thus, unless the line-of-sight projection \refeq{lambda} can somehow be 
restricted to a range $\D\chi_\lambda \lesssim D^2(\bar z) L_p \approx 54.6\, D^2(\bar z)\Mpch$, the impact
of primordial non-Gaussianity on the clustering of peaks identified in a 
projected density field is strongly suppressed.  

\section{Conclusion}
\label{sec:disc}

Observations of the large-scale structure in the Universe offer
very promising opportunities for probing the initial conditions of the
Universe, complementing the searches for deviations from Gaussianity
in the temperature anisotropies of the cosmic microwave background.
Weak lensing is one of the most powerful probes of large-scale 
structure, as it directly probes the underlying matter distribution, thus
circumventing many of the uncertainties in the tracer-mass relation.  

In this paper, we have shown that shear and convergence power spectra are in 
principle subject to corrections from primordial non-Gaussianity, since 
weak lensing estimators generally have non-linear corrections.  However,
projection effects severely reduce the impact of primordial non-Gaussianity,
and the effects are generally much smaller than those from nonlinear
gravitational evolution.  This also holds when approximately
including the effect of non-linear growth on small scales.  We also 
investigate the same effects for galaxy-galaxy lensing, {\em i.e.} the cross-correlation of shear and convergence with some large-scale structure tracer.  In this case, the suppression is somewhat mitigated, but still significant.  

Finally, in \refsec{peaks} we provide a general argument why this suppression
is generic to searching for non-Gaussianity in two-dimensional projected
density fields.  If the projection occurs over a longer line-of-sight distance 
than the one-dimensional coherence length of the density field, 
$L_P \approx 54.6\:\Mpch$ (at $z=0$),  
the projected density field is more Gaussian than the underlying 3D density
field, a consequence of the central limit theorem \cite{scoccimarro/zaldarriaga/hui:1999}.  
Thus, the effect
of primordial non-Gaussianity on the clustering of peaks identified in
such a projected density field is suppressed by $\sim L_P/\D\chi$,
where $\D\chi$ is the width of the projection kernel.  This
applies to peaks identified in weak lensing shear maps (where $\D\chi\sim H_0$)
as well as in diffuse backgrounds---unless a definite connection between such
peaks and dark matter halos in the 3D density field can be made.  
Thus, we expect that the statistics of tracers of the 3D density field will 
generally provide tighter constraints on primordial non-Gaussianity
than those of projected fields.

\acknowledgments
We would like to thank Francis Bernardeau, Duncan Hanson, Chris Hirata, Toshiya Namikawa, and 
Atsushi Taruya for enlightening discussions.  DJ and FS are supported
by the Gordon and Betty Moore Foundation at Caltech. ES acknowledges support by the European Commission under the Marie Curie Inter European Fellowship. We are grateful to the organizers of the ``Cosmo/CosPA 2010'' conference at the University of Tokyo, Japan, where this work was initiated.



\end{document}